\documentclass[sigconf,pbalance]{acmart}

\copyrightyear{2027}
\acmYear{2027}
\setcopyright{cc}
\setcctype{by}
\acmConference[WSDM '27]{Proceedings of the Twentieth ACM International Conference on Web Search and Data Mining}{February 15--19, 2027}{Hong Kong, Hong Kong}
\acmBooktitle{Proceedings of the Twentieth ACM International Conference on Web Search and Data Mining (WSDM '27), February 15--19, 2027, Hong Kong, Hong Kong}
\acmDOI{10.1145/3862322.3863570}
\acmISBN{979-8-4007-3117-4/2027/02}

\usepackage{array}
\usepackage{multirow}

\graphicspath{{figures/}}

\newcommand{\tool}{FocalSwap}
\newcommand{\hpre}{H_{\mathrm{pre}}}

\newcommand{\ypost}{Y_{\mathrm{post}}}
\newcolumntype{L}[1]{>{\raggedright\arraybackslash}p{#1}}

\begin{document}

\title{Do Search-Augmented Recommenders Use the Latest Search Episode?}

\author{Jiandong Ding}
\orcid{0000-0001-8123-9889}
\affiliation{%
  \institution{Huawei Technologies Co., Ltd.}
  \city{Shanghai}
  \country{China}
}
\email{dingjiandong2@huawei.com}
\renewcommand{\shortauthors}{Jiandong Ding}

\begin{abstract}
When a user searches and then returns to a recommendation feed, does the
latest search episode actually help rank the next item? While search-augmented
recommenders (SAR) report overall ranking gains, aggregate metrics leave open
whether these gains stem from prediction-time use of the latest search or from
broader search state and representation learning. To isolate prediction-time
search attribution, we introduce \mbox{\textbf{\tool}}, a counterfactual audit
that masks or replaces either the latest search episode or the full search
history while holding the target, candidates, recommendation history, and
trained scorer strictly fixed. We first verify that \tool{} reliably detects
known search dependence in a planted-signal test and in GenSAR's native
search-item prediction task. In the proxy analyses, KuaiSAR's routing gain
falls below the shuffled threshold, and Qilin shows its clearest
matched-search advantage at high taxonomy overlap.

Auditing trained SESRec and UniSAR checkpoints reveals that primary
comparisons do not establish a positive ranking gain from the latest search
episode alone. For the full search history, attribution depends critically on
the reference control. On a shared set of 66{,}139 instances across three
UniSAR checkpoints, neutral masking consistently lowers mean reciprocal
rank (RR). However, replacing the search history with observed searches
matched on category and activity profile alters nearly half the target ranks
(44.1\%--44.8\%) yet leaves the original history with little average
advantage over these replacements. Across checkpoints, matched searches
under the primary assignment yield a mean RR $2.770\times10^{-3}$ to
$6.604\times10^{-3}$ higher than neutral masking. The adjusted confidence
intervals lie above zero for the primary assignment and all five alternative
assignments per checkpoint. Because scorer sensitivity to input removal does
not imply content-specific benefit, search audits should explicitly report
both how often ranks move and how much the original search helps, specifying
which search inputs changed and what replaced them.
\end{abstract}



\maketitle

\section{Introduction}

When someone searches for an item and then opens a recommendation feed,
the search they just made seems like an obvious clue to what they want next.
Yet the ranking gains of search-augmented recommendation (SAR) leave open
where those gains come from \cite{si2023sesrec,xie2024unifiedssr,shi2024unisar}.
Search can help train item and user representations, earlier searches can
shape the model's view of the user's interests, and the latest episode can
provide a clue at prediction time. The question arises across architectures,
from joint graph learning \cite{zhao2022srjgraph} and intent transfer
\cite{zhang2024uditsr} to generative retrieval \cite{shi2025gensar}.

To see what the latest episode contributes at prediction time, we change
only that episode and ask how the same recommendation would rank.
The target, candidates, recommendation history, and trained scorer all stay
fixed. Changing the entire search history asks a broader question, and
in either case the replacement matters. A neutral mask removes search
content and can introduce an unfamiliar input. A replacement from real
search logs preserves selected properties, such as category and search
profile, while supplying different content.

\begin{figure*}[!t]
  \centering
  \includegraphics[width=\textwidth]{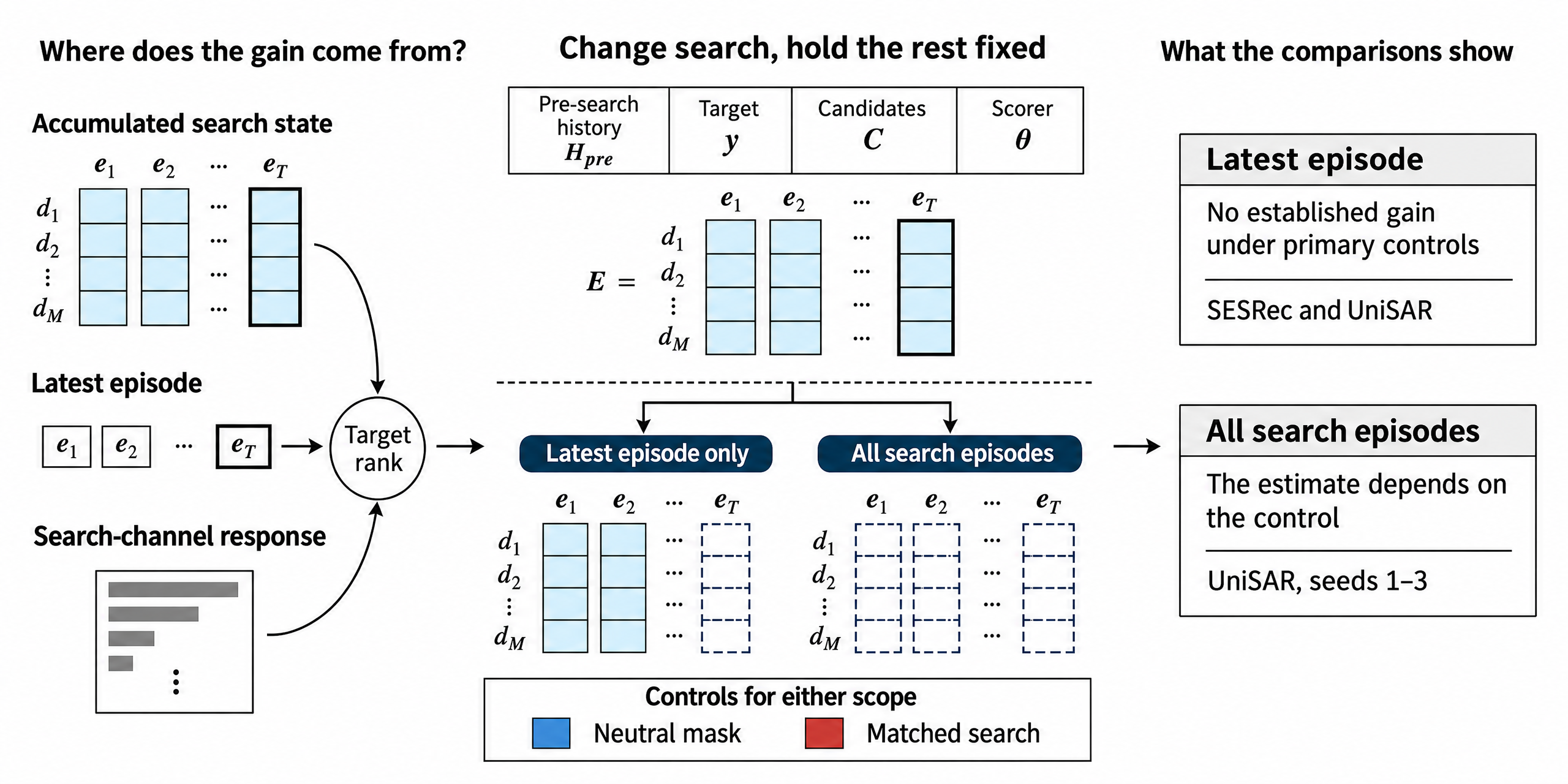}
  \caption{\tool{} fixes the recommendation history (shown as $\hpre$ for
  a search-centered window), target, candidates, and scorer. It masks or replaces either the most
  recent episode or the full search history, as indicated by the dashed
  columns. The primary comparisons do not establish a latest-episode gain
  in SESRec or UniSAR, and the full-history estimate in UniSAR depends on
  the control.}
  \Description{Three panels show why a change in target rank can have multiple
  explanations, then illustrate the fixed inputs and two intervention scopes
  using search-slot matrices. Dashed columns identify which episodes change.
  Neutral masking and matched observed search are controls for either scope.
  The final panel summarizes the primary latest-episode results for SESRec
  and UniSAR and the full-history control comparison for UniSAR.}
  \label{fig:overview}
\end{figure*}

\tool{} makes these comparisons for each evaluation instance
(Figure~\ref{fig:overview}). It measures how often the target moves,
how much the original search improves reciprocal rank (RR) relative to
a control, and how often the original search wins against a matched
alternative, giving ties half credit. We call the episode closest to the
recommendation the \emph{focal search}.

The primary comparisons in SESRec and UniSAR leave a ranking gain from
the latest episode unestablished (Figure~\ref{fig:checkpoint-control-sensitivity}(a)).
For the full search history in UniSAR, the choice of control changes the
answer. On the same 66{,}139 instances, masking yields a positive
contribution at all three checkpoints. Matched replacements yield smaller
estimates, even though they change the target rank in nearly half the
instances. The scorer responds to the search input, and the comparison
determines how much benefit we attribute to the original searches.

KuaiSAR supplies same-user search and recommendation histories with
anonymous tokens \cite{sun2023kuaisar}, and Qilin provides real query and
item text \cite{chen2025qilin}. The behavioral and lexical analyses ask
when history and search offer complementary ranking signals. A planted
signal and GenSAR's native query task check that the audit can detect
dependence on a particular search.

Our contributions are to distinguish the latest search episode from the
accumulated search history as possible sources of ranking gains, introduce
an audit that detects known search signals, and show how the reference
input changes attribution in trained recommenders. Table~\ref{tab:evidence-map}
previews the results. For each audit, we report what changed, what replaced
it, and how the ranking responded.

\begin{table*}[!t]
\centering
\caption{Where each evaluation contributes evidence.}
\label{tab:evidence-map}
\small
\setlength{\tabcolsep}{3.2pt}
\renewcommand{\arraystretch}{1.05}
\begin{tabular*}{\textwidth}{@{\extracolsep{\fill}}L{0.20\textwidth}L{0.17\textwidth}L{0.59\textwidth}@{}}
\toprule
Setting & Question & Finding \\
\midrule
SESRec; UniSAR seeds 1--3 & Latest episode & The primary comparisons do not establish a positive ranking contribution. \\
UniSAR seeds 1--3 & Full history & Masking yields a positive contribution on 66,139 shared instances. Matched replacement gives smaller estimates. \\
Planted signal & Calibration & The audit detects routing opportunity, contribution, and specificity at full signal strength in all three runs. \\
GenSAR & Calibration & The original query raises mean RR@30 over matched replacement on 8,958 search-item instances. \\
KuaiSAR and Qilin & Auxiliary analyses & KuaiSAR falls below the shuffled threshold. Qilin clears the 95th-percentile threshold with BM25. Its strongest matched-search result is a $0.084$ preference lift on 6.1\% of instances. \\
\bottomrule
\end{tabular*}
\end{table*}

\section{Related Work}

We relate SAR models, attribution audits, and search-rich datasets to
distinguish system-level utility from evidence about the latest search episode.

\paragraph{Search-augmented and joint search--recommendation models.}
SAR builds on the broader implicit-feedback and sequential recommendation
line, where ranking objectives and sequence encoders such as BPR, SASRec, and
BERT4Rec define standard recommendation backbones
\cite{rendle2012bpr,kang2018sasrec,sun2019bert4rec}.
HPGR models session structure and attends selectively to preference-relevant
history \cite{chen2026hpgr}.
Early joint systems model search and recommendation behavior together
\cite{zamani2020joint,zhao2022srjgraph,bhattacharya2024unicorn}.
Closer to our checkpoint analysis, SESRec learns a search representation for
recommendation \cite{si2023sesrec}, UnifiedSSR unifies sequential search and
recommendation \cite{xie2024unifiedssr}, and UniSAR models transitions between
the two \cite{shi2024unisar}. In parallel, work on shared representations
links search and recommendation by modeling integrated behavior,
transferring intent, or disentangling signals
\cite{yao2021user,zhang2024uditsr,cui2024clardrec,shi2025lcrser}.
This line also addresses sparse search histories \cite{shi2025gserec},
different interest horizons \cite{zhu2026dhim}, and multiple recommendation
scenarios \cite{liu2024usr,xia2025superrs}. Generative models bring the two
tasks together through identifiers, intent, and queries
\cite{liao2026uimg,shi2025gensar,zhao2025gensr,yan2025intsr}, shared training
\cite{penha2024bridging,zhao2026gems}, or a unified task formulation
\cite{denadai2026neo,guo2026onefeed}.

Across these designs, search can shape representations during training,
accumulate in user state, or enter the scorer at prediction time. \tool{}
compares the search evidence attached to one recommendation instance: changing the
latest episode tests focal attribution, while changing all search inputs tests
broader search-state contribution. GenSAR's native current-query task provides
a separate calibration of known query use \cite{shi2025gensar}.

\paragraph{Evaluation audits and signal attribution.}
Recommendation evaluation has a long history of re-examining whether reported
gains survive stronger baselines, controlled comparisons, or reproducibility
checks \cite{ferrari2019progress,dacrema2019troubling,shehzad2025worrying}.
Behavioral testing systems make a related distinction between aggregate metrics
and targeted capability checks \cite{ribeiro2020checklist,chia2021reclist}.
Diagnostic resources similarly examine components behind ranking outcomes:
SIDInspector evaluates Semantic-ID mappings separately from downstream scores
\cite{ding2026sidinspector}, while SIDScope traces their effects on candidate
retrieval and item resolution \cite{ding2026sidscope}.
Model-interpretability research exposes a complementary risk in input
attribution. Amnesic probing distinguishes information encoded in a
representation from information used by the model through a behavioral
intervention \cite{elazar2021amnesic}. Feature-attribution studies further show
that the baseline chosen to represent missingness can change an explanation and
that test-time feature removal can create out-of-distribution inputs
\cite{sturmfels2020visualizing,hase2021ood}. \tool{} brings these concerns to
temporal SAR attribution: it fixes the complete recommendation instance, separates
latest-episode from accumulated-search claims, and compares neutral masking
with matched observed search sequences on identical instances.
A separate line of research uses logged feedback to estimate policy value
\cite{bottou2013counterfactual,dudik2014doubly,swaminathan2015crm,gilotte2018offline}
or correct for exposure and selection bias
\cite{schnabel2016treatments,joachims2016unbiased,saito2019unbiased}.
Related causal methods target particular sources of confounding. DIFL learns
duration-invariant features for watch-time prediction \cite{jin2026invariant},
while IV4Rec uses search behavior as an instrument to identify causal
preferences \cite{si2022iv4rec}. In \tool{}, the question is how a fixed
scorer ranks the same target when given a different search.

\paragraph{Intent, transfer, and search-rich datasets.}
Search-rich datasets expose intent through behavior tokens, query text, or
richer semantic representations. KuaiSAR provides same-user search and recommendation behavior
but anonymizes tokens \cite{sun2023kuaisar}. JDSearch and KuaiSearch expose
real-query-derived tokens and retrieval context \cite{liu2023jdsearch,
li2026kuaisearch}, and query representation learning offers another path for
using explicit search signals in recommendation \cite{wang2024queryctr}. Qilin
adds real query and item text in a search/recommendation behavior setting
\cite{chen2025qilin}. The available inputs determine which aspects of search each dataset can test.

\section{FocalSwap Protocol}

Consider a validation instance with a planted search signal
(Section~\ref{sec:planted-signal}). The target ranks first with the
original search and recommendation history
($\mathrm{RR}=1.000$). It ranks third with either a neutral mask or a matched
search ($\mathrm{RR}=0.333$). With the same candidates and scorer in every
view, the audit records a rank change, a positive contribution, and a win
over the replacement. History alone gives RR of 0.333, while search alone
gives 1.000.

We link search windows to later recommendations and report the coverage
of those links. An eligible instance $r_i=(H_i,E_i,Y_i,C_i)$ has the scorer's
recommendation history $H_i$, ordered search input $E_i$, and candidates
$C_i$. Its nonempty evaluated positives satisfy
$Y_i\subseteq Y_{\mathrm{post},i}\cap C_i$, where $\ypost$ denotes the
post-search positives. The latest non-padding episode in $E_i$ is focal.
Under view $V$, let $R_i(V)$ be the best positive rank (starting at 1):
\begin{equation}
  R_i(V)=\min_{y\in Y_i}\operatorname{rank}(y;s_\theta(V_i,C_i)),
  \qquad U_i(V)=R_i(V)^{-1}.
  \label{eq:fixed-row-utility}
\end{equation}
The target and candidate sets stay fixed across views. SESRec and UniSAR
use midpoint ranks for exact candidate-score ties. Eligibility also requires
that the target cannot leak through the history or focal search fields
supplied to the scorer.

$F$ combines history and the original search, $H$ uses history alone,
and $S$ uses the focal search alone. Intervention $m$ maps $F_i$ to
$M_{m,i}$, with $\widetilde F$ denoting matched replacement. We average
over the nonempty set $\Omega_m$ where $m$ applies. Matching requires a
valid replacement.

We put a neutral ID or matched observed content in either the latest
non-padding slot or all non-padding search slots. This measures how the
fixed scorer reranks the same candidates.

\subsection{What the Audit Measures}

The intervention analysis tracks three quantities. The behavioral and
lexical settings also let us ask whether one can improve ranking by
routing between history-only and search-only predictions.

First, rank response measures how often the best positive rank changes.
The indicator $\mathbf{1}$ equals one when the ranks differ:
\begin{equation}
  \widehat\rho_{\mathrm{rank}}(m)
  =\frac{1}{|\Omega_m|}\sum_{i\in\Omega_m}
  \mathbf{1}\!\left[
  R_i(F)\ne R_i(M_m)\right].
  \label{eq:rank-response}
\end{equation}
Second, signed ranking contribution measures whether the original search input
improves average RR relative to the named control:
\begin{equation}
  \widehat{\Delta}_{\mathrm{contrib}}(m)
  =\frac{1}{|\Omega_m|}\sum_{i\in\Omega_m}
  \left(U_i(F)-U_i(M_m)\right).
  \label{eq:contribution}
\end{equation}
Frequent positive and negative rank changes can yield a small signed mean,
so response rate and contribution answer different questions. On the same
eligible instances, subtracting two contribution estimates cancels the
original RR. The paired gap is the mean RR under the second control minus
the first.

Third, replacement preference compares $V$ to its match $\widetilde V$.
Checkpoint swaps compare complete inputs ($V=F$), and proxy swaps compare
search-only predictions ($V=S$). The comparison uses RR for $A_i(V)$.
The shared-candidate diagnostic uses the target score. For $N$ matched
instances, ties receive half credit:
\begin{equation}
  z_i=
  \begin{cases}
    1, & A_i(V)>A_i(\widetilde V),\\
    \tfrac{1}{2}, & A_i(V)=A_i(\widetilde V),\\
    0, & A_i(V)<A_i(\widetilde V),
  \end{cases}
  \qquad
  \widehat h_{\mathrm{swap}}=\frac{1}{N}\sum_{i=1}^{N}z_i.
  \label{eq:half-tie}
\end{equation}
A tie-adjusted preference of 0.500 means that wins and losses balance.
Table~\ref{tab:checkpoint-summary}(b) reports the statistic for replacing
the latest episode or the full history. Evidence for focal attribution
requires both positive signed contribution and a preference for the
original over matched replacements under the calibration and uncertainty
procedures below.

\paragraph{Routing with history and search.}
In the behavioral and lexical settings, we can also ask whether history and
search make complementary predictions.

A per-instance oracle selects whichever of History-only and Search-only yields
higher RR. Its margin over History-only bounds the gain from routing between those
views, but selecting the better score can create headroom even after
shuffling. A shuffled-search null calibrates this selection effect. The observed margin is
\begin{equation}
  \widehat{\Delta}_{\mathrm{opp}}
  =\frac{1}{N}\sum_{i=1}^{N}
  \left(\max\{U_i(H),U_i(S)\}-U_i(H)\right).
  \label{eq:opportunity}
\end{equation}
In each of $B$ null draws, we hold the history predictions fixed and
shuffle or replace the search predictions. Let
$\widehat{\Delta}_{\mathrm{opp}}^{\pi_b}$ be the same margin in draw $b$.
Routing opportunity clears the 95th-percentile null threshold when
\begin{equation}
  \widehat{\Delta}_{\mathrm{opp}}
  >\widehat q_{0.95}
  \left(\left\{\widehat{\Delta}_{\mathrm{opp}}^{\pi_b}\right\}_{b=1}^{B}\right).
  \label{eq:opportunity-gate}
\end{equation}
The recovery probe measures how much of that oracle margin a routing rule
recovers on held-out instances using features that exclude the target outcome.

Checkpoint intervals and Qilin swap intervals resample whole users.
The KuaiSAR swap intervals resample instances. In Qilin, we apply Holm
correction to one-sided user-cluster sign-flip tests across the ten swap buckets.
For each primary checkpoint endpoint (matched contribution and the
paired gap), Bonferroni-adjusted 97.5\% intervals cover its two scopes.

\paragraph{Choosing the reference input.}
A neutral ID removes search content and may be rare or absent in the
scorer's usual inputs. We therefore also use replacements that preserve
selected category or profile properties. The same-scorer comparisons
apply to proxy scorers, controlled interfaces, and trained checkpoints.
For released implementations, the native task and performance determine
which comparisons enter the checkpoint analysis.

\section{Evaluation Settings}

\subsection{KuaiSAR Behavioral Evidence}

We construct search-centered KuaiSAR windows from anonymized search and
recommendation interactions, linking pre-search recommendation history, focal
search evidence, and post-search positives. We remove targets from the
scored history and search fields. The fixed candidates include the
post-search positives, bounded history and search items, and train-split
popularity negatives, capped at 80 (mean 48.8 on held-out instances).

A deterministic ranker scores the history, original search, and replacement
search using transitions learned on the training split. It counts distinct evidence--target pairs separately for the pre-search
history and focal search in each training instance. Let \(c(e,t)\) count transitions
from evidence item \(e\) to target \(t\), and let
\(\mathrm{total}(e)=\sum_t c(e,t)\). For evidence set \(E\) and candidate
\(t\), it scores
\begin{equation}
  s(E,t)=\max_{e\in \mathrm{uniq}(E)}
  \frac{\log(1+c(e,t))}{\log(2+\mathrm{total}(e))}.
  \label{eq:proxy-score}
\end{equation}
The maximum selects the strongest evidence-to-target association, and the
denominator reduces the weight of frequent evidence items. Unseen pairs and
empty evidence score zero. Equal scores use deterministic hash tie-breaking.

KuaiSAR analyzed instances are selected deterministically by sorting eligible
examples by \texttt{MD5(example\_id)} within each split and taking 10{,}000
validation and 10{,}000 test instances.

\subsection{Qilin Lexical Evidence}

Qilin ranks each logged recommendation slate with BM25 and taxonomy
overlap. The search branch uses the query and clicked-item taxonomy paths
from that search, both of which change with a replacement. The history
branch builds a BM25 pseudo-query and taxonomy features from up to 20 clicks
in the recommendation request's history, which can include focal-search clicks.
For each distinct path, matching nonempty levels add 0.04 to the search
score or 0.03 to the history score. RR uses the first ranked clicked item.
Score ties follow the logged slate order.

The Qilin construction records both match coverage and match quality. Same-user
and taxonomy-near swaps are available for nearly all instances. For taxonomy matching,
we count distinct taxonomy paths among the donor's clicked items whose top-level
category also appears among the focal search's clicked items. The bucket
tax$\geq k$ contains instances with at least \(k\) such donor paths. A random-search
placebo gives each bucket a reference on the same instances. This setup lets the
audit distinguish a broad lexical effect from a local taxonomy-overlap effect
on fixed targets and candidates.

\begin{figure*}[!t]
  \centering
  \includegraphics[width=\textwidth]{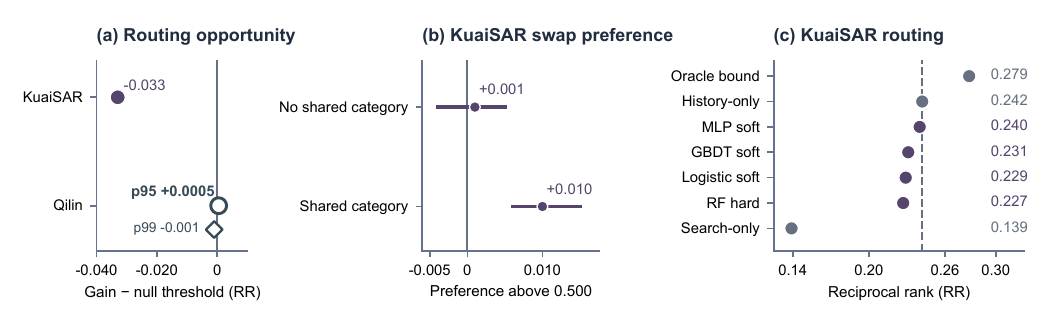}
  \caption{Search and history in the proxy evaluations. Panel (a) compares
  observed oracle gains with the shuffled or counterfactual null thresholds.
  Panel (b) shows KuaiSAR's tie-adjusted preference above 0.500 for
  same-source replacements, with 95\% instance-bootstrap intervals.
  Panel (c) compares KuaiSAR routers with the history baseline and oracle.}
  \Description{Three panels show the routing opportunity in KuaiSAR and Qilin,
  then the KuaiSAR swap preference and KuaiSAR recovery scores. The Qilin
  point is above the 95th and below the 99th percentile null threshold. The
  KuaiSAR routers fall below the history-only baseline.}
  \label{fig:gate-evidence}
\end{figure*}

Table~\ref{tab:audit-scale} shows how many episodes enter the proxy
analyses after the fixed-instance eligibility checks.
Figure~\ref{fig:gate-evidence} compares the two settings, and
Figure~\ref{fig:swap-specificity} shows the Qilin replacement results by
matching rule.

\begin{table}[t]
\centering
\caption{Behavioral and lexical evaluation scale. Valid episodes satisfy
the fixed-instance criteria; analyzed episodes enter the proxy analyses.}
\label{tab:audit-scale}
\small
\begin{tabular*}{\columnwidth}{@{\extracolsep{\fill}}lrrr@{}}
\toprule
Dataset & Users & Valid episodes & Proxy episodes \\
\midrule
KuaiSAR & 13,159 & 96,994 & 20,000 \\
Qilin   & 1,544  & 6,135  & 6,135 \\
\bottomrule
\end{tabular*}
\end{table}

\subsection{Counterfactuals and Calibration}

For KuaiSAR, nuisance matching (L1) picks a search from the same split
and source, with similar feedback count, pre-history length, and bridge length.
Category matching (L2) samples a search from the same split that shares
at least one search category. In Qilin, we use same-user, time-near, and
taxonomy-based replacements, and compare them with random search on the
same instances. The Qilin routing null draws uniformly from these four
replacement predictions for each instance. Recovery is measured against
History-only within the same candidate pool. Qilin's router is an offline
diagnostic using features from the logged replacement pool.

\subsection{Checkpoints from Released Code}

The recommendation audits use SESRec and UniSAR checkpoints trained with
the released implementations on KuaiSAR-compatible inputs. UniSAR seeds
1--3 provide three independent checkpoints tested on 157{,}192 recommendation
instances. SESRec uses 16{,}788 eligible test windows. Each model ranks
one target against 99 model-specific negatives, fixed across interventions.
The recommendation context stays as supplied to the checkpoints, including
interactions after the focal search when present. Full-input RR is 0.0423 for SESRec and 0.3417, 0.3542, and 0.3403 for the
three UniSAR seeds. The KuaiSAR proxy uses the separate candidate sets capped at 80 items.

SESRec takes one search session as input, so intervening on it also changes
the full search input. Its replacement shares at least one search-item
category. UniSAR's matched replacement also uses search-profile buckets.
GenSAR's recommendation checkpoint scored 0.1857 in test NDCG@10, below
the prespecified inclusion threshold of 0.1914 (95\% of the published
score). We use its native search-item prediction task (PerSrc) to check
whether the audit detects use of the current query.

\section{Positive Calibration}
\label{sec:positive-control}

Before auditing recommendation checkpoints, we verify that \tool{} detects
input dependence when it is known to be present. One score construction plants
a focal signal; a released scorer explicitly conditions its native task on the
current query. Together they calibrate the recommendation-side analysis.

\subsection{Planted Focal Signal}
\label{sec:planted-signal}

An auxiliary score-sensitivity check adds a boost only when the focal
search has positive feedback and does not directly conflict with pre-search history. With no
injection, oracle gain is 0 and the injected-versus-passive preference is 0.500.
At the smallest positive signal strength (oracle gain 0.003), preference
reaches 0.645 and a held-out two-layer MLP router recovers the injected gain. At the largest
signal strength, oracle gain reaches 0.042 and recovery persists. This check isolates
score sensitivity and routing. The following full-chain control adds matched
donors and the shuffled-search opportunity null.

The separate full-chain control keeps Qilin inputs, candidates, and donors
fixed and generates synthetic target labels with a fixed teacher. The teacher
standardizes history and focal-search scores within each candidate set, then
adds the history score, the weighted focal-search score, and fixed noise.
The highest-scoring candidate becomes the target and stays fixed across all
audit views. For each injection weight (0, 0.5, or 1.0), we train the student scorer with
three seeds. With zero signal weight, Full-minus-masked RR ranges from $-0.0070$ to $-0.0041$:
neutral masking changes ranking even without a planted focal signal. None of these zero-signal runs passes the positive contribution gate. At full signal weight (1), the signed RR contribution under masking is
0.052--0.054, preference for the original search is 0.639--0.651, and
placebo-adjusted specificity is 0.147--0.163; every interval clears its null
boundary. The planted control therefore establishes routing opportunity,
signed contribution, and focal specificity for all three full-strength runs.
On its test split, the full-strength run used for the worked example gives
an oracle gain of 0.128 against a shuffled p95 of 0.068, a contribution of
$+0.054$ RR, and a tie-adjusted preference of 0.647.

\subsection{Released-Code Current-Query Signal}
\label{sec:gensar-calibration}

GenSAR provides a second positive calibration on its native search-item
prediction task \cite{shi2025gensar}. On 8{,}958 leakage-filtered validation
instances, RR@30 uses the 30 beam results and assigns zero when they omit
the target. Masking the current query reduces RR@30 by 0.035
(95\% CI [0.029, 0.041]). A matched query reduces it by 0.106
(95\% CI [0.099, 0.112]), with a preference of 0.635 for the original.
Replacing only the most recent prior search session reduces RR@30 by 0.002
and gives a preference of 0.518. These results recover current-query
dependence at the preselected epoch-3 checkpoint.

\begin{figure*}[!t]
  \centering
  \includegraphics[width=\textwidth]{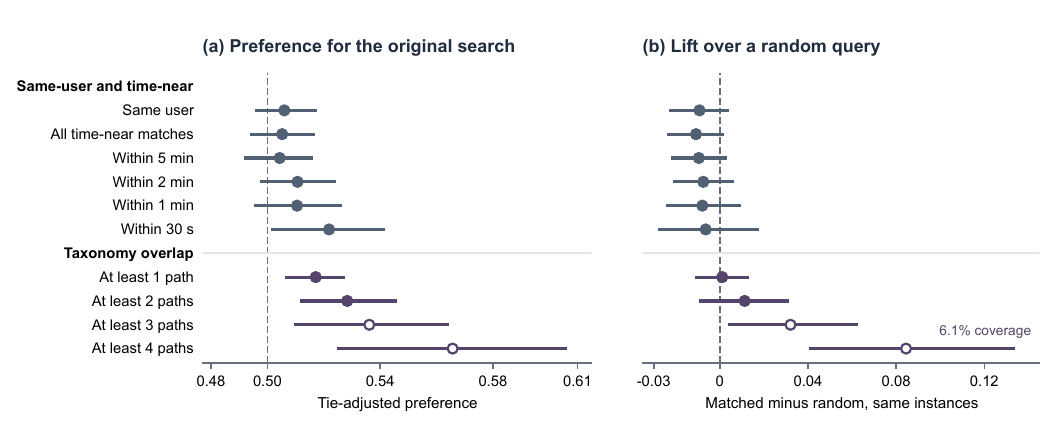}
  \caption{Qilin's matched-search signal concentrates at high taxonomy overlap.
  Panel (a) gives the tie-adjusted preference for the original search. Panel
  (b) subtracts the preference against a random search on the same instances.
  Bars are pointwise 95\% confidence intervals. Open markers denote below
  20\% coverage. Only the at-least-four-path bucket passes Holm correction.}
  \Description{Two aligned plots show preference for the original search and
  its lift over a random-query replacement. Rows group the same-user and
  time-near comparisons separately from the taxonomy-overlap comparisons.
  The last row covers 6.1 percent of instances and gives a lift of 0.084.}
  \label{fig:swap-specificity}
\end{figure*}

\begin{figure*}[!t]
  \centering
  \includegraphics[width=\textwidth]{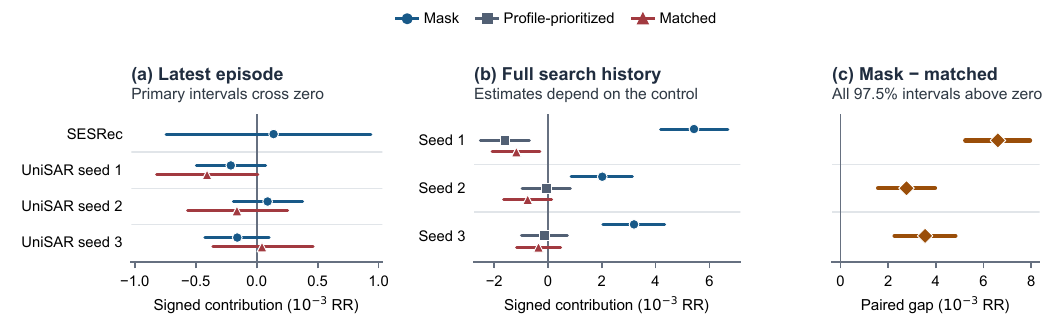}
  \caption{Search contribution under the original replacement assignment.
  In (a), the intervals cross zero for masking (16,788 SESRec and 157,192
  UniSAR instances) and UniSAR matched replacement (87,044 instances).
  In (b) and (c), all controls use the same 66,139 UniSAR instances.
  Masking gives positive estimates, and the paired mask-minus-matched
  intervals are above zero for all three seeds. User-cluster intervals
  are 95\% for masking and profile-prioritized replacement, and
  Bonferroni-adjusted 97.5\% for matched replacement and the paired gaps.}
  \Description{Three forest plots show signed RR estimates for the latest
  search episode, the full search history, and the paired full-history gap,
  with a line at zero in each panel. All primary latest-episode intervals
  cross zero. For the full search history, masking gives higher contribution
  estimates than either observed replacement. The three paired intervals
  lie above zero.}
  \label{fig:checkpoint-control-sensitivity}
\end{figure*}

\section{Behavioral and Lexical Evidence}

The proxy analyses ask where search provides a complementary ranking
signal. History-only and search-only predictions let us test whether
there is room to improve by routing between them and how much of that
improvement a router can recover (Figure~\ref{fig:gate-evidence}).

\subsection{Coverage and Opportunity Separate}

KuaiSAR offers broad episode coverage under the target-leakage checks,
but the gain from choosing the better prediction is 0.037, below the
shuffled-search 95th percentile of 0.070. The proxy falls short of the
routing-opportunity threshold.

Qilin's BM25 scorer clears its 95th-percentile null threshold by 0.0005 RR
($p=0.030$).
Across 30 Monte Carlo null seeds, the observed oracle gain exceeds the
95th percentile in each run. It stays below the pooled 99th percentile,
and the result does not replicate with TF-IDF. The opportunity is small
and specific to BM25.

\begin{table*}[t]
\centering
\caption{Recommendation checkpoint results. Panel (a) uses the original assignment.
Panel (b) reports tie-adjusted preference (C2 for UniSAR, L2 for SESRec)
and the profile-prioritized contribution. Panel (c) varies the matched
assignment. RR contributions and gaps are in $10^{-3}$ units. Brackets
give user-cluster intervals. Bold contribution intervals exclude zero.
SESRec takes one search session and uses a category-only match. Its
matched contribution is reported in the text, and its preference appears
once in (b).}
\label{tab:checkpoint-summary}
\small
\setlength{\tabcolsep}{2.4pt}
\renewcommand{\arraystretch}{1.12}
\begin{tabular*}{\textwidth}{@{\extracolsep{\fill}}l c c c c c@{}}
\toprule
\multicolumn{6}{l}{(a) Signed contribution under masking and matched replacement} \\
& \multicolumn{2}{c}{Latest episode} & \multicolumn{3}{c}{Full history on 66{,}139 shared instances} \\
\cmidrule(lr){2-3}\cmidrule(lr){4-6}
Checkpoint & Mask (C0), 95\% & Matched (C2), 97.5\% & Mask, 95\% & Matched, 97.5\% & Mask $-$ matched, 97.5\% \\
\midrule
SESRec & $+0.138\;[-0.743,+0.936]$ & n/a & n/a & n/a & n/a \\
UniSAR seed 1 & $-0.213\;[-0.495,+0.072]$ & $-0.408\;[-0.822,+0.009]$ & \textbf{\boldmath $+5.427\;[+4.183,+6.689]$} & \textbf{\boldmath $-1.177\;[-2.074,-0.302]$} & \textbf{\boldmath $+6.604\;[+5.238,+7.951]$} \\
UniSAR seed 2 & $+0.089\;[-0.191,+0.376]$ & $-0.163\;[-0.567,+0.253]$ & \textbf{\boldmath $+2.016\;[+0.856,+3.138]$} & $-0.755\;[-1.647,+0.131]$ & \textbf{\boldmath $+2.770\;[+1.582,+3.962]$} \\
UniSAR seed 3 & $-0.160\;[-0.426,+0.102]$ & $+0.043\;[-0.361,+0.462]$ & \textbf{\boldmath $+3.197\;[+2.040,+4.331]$} & $-0.354\;[-1.171,+0.469]$ & \textbf{\boldmath $+3.551\;[+2.272,+4.819]$} \\
\bottomrule
\end{tabular*}
\par\medskip
\begin{tabular*}{\textwidth}{@{\extracolsep{\fill}}l c c c@{}}
\toprule
\multicolumn{4}{l}{(b) Replacement preference and supplementary contribution (pointwise 95\% intervals)} \\
Checkpoint & Latest episode $\widehat h_{\mathrm{swap}}$ & Full history $\widehat h_{\mathrm{swap}}$ & Profile-prioritized (C1), full history \\
\midrule
SESRec (L2) & $0.4966\;[0.4868,0.5065]$ & n/a & n/a \\
UniSAR seed 1 & $0.4977\;[0.4960,0.4994]$ & $0.4899\;[0.4870,0.4928]$ & \textbf{\boldmath $-1.600\;[-2.516,-0.691]$} \\
UniSAR seed 2 & $0.4982\;[0.4965,0.4999]$ & $0.4959\;[0.4929,0.4989]$ & $-0.068\;[-0.959,+0.830]$ \\
UniSAR seed 3 & $0.4993\;[0.4975,0.5013]$ & $0.4957\;[0.4928,0.4986]$ & $-0.132\;[-0.986,+0.722]$ \\
\bottomrule
\end{tabular*}

\par\medskip
\begin{tabular*}{\textwidth}{@{\extracolsep{\fill}}l c c c@{}}
\toprule
\multicolumn{4}{l}{(c) Full history across five alternative matched assignments} \\
Checkpoint & Matched range & Mask $-$ matched range & Paired 97.5\% CIs above zero \\
\midrule
UniSAR seed 1 & $-1.679$ to $-0.953$ & $+6.380$ to $+7.106$ & $5/5$ \\
UniSAR seed 2 & $-0.837$ to $-0.316$ & $+2.331$ to $+2.852$ & $5/5$ \\
UniSAR seed 3 & $-0.534$ to $+0.141$ & $+3.057$ to $+3.732$ & $5/5$ \\
\bottomrule
\end{tabular*}

\end{table*}

\subsection{Matched-Swap Signal Is Local}

Within the same split and data source in KuaiSAR, the tie-adjusted
preference is 0.501 for replacements with no shared category and 0.510
for replacements with a shared category. The latter is $+0.010$ over
symmetry, a small advantage on that comparison.

In Qilin we compare each taxonomy-overlap bucket with random-search
replacement on the same instances (Figure~\ref{fig:swap-specificity}).
Only the strictest bucket, with at least four qualifying paths, passes Holm correction.
It covers 6.1\% of instances, with a tie-adjusted preference of 0.566
and a lift of $+0.084$ over random replacement ($p_{\mathrm{Holm}}=0.003$).

KuaiSAR's routers do not improve on history alone. Qilin's small BM25
opportunity also produces no held-out routing gain, including in the
at-least-four-path bucket.

\section{Checkpoint Evidence}
\label{sec:checkpoint-readouts}

Figure~\ref{fig:checkpoint-control-sensitivity} shows what happens when
we change the latest episode and when we change the full search history.
Table~\ref{tab:checkpoint-summary} gives the estimates and interval endpoints.

\paragraph{What replaces the search.}
The mask (C0) puts a checkpoint-compatible neutral ID in the selected
slots. The profile-prioritized control (C1) prefers similar item counts,
query lengths, and frequencies, with broader fallback pools and no category
requirement. The matched
control (C2) uses both category and profile. In UniSAR, a matched search
is available for 87{,}044 of 157{,}192 instances (55.4\%) at the latest
slot. Replacing the full history requires a match at every slot, which
leaves 66{,}139 instances (42.1\%). For that comparison, we apply all
three controls to these same instances.

\subsection{The Latest Episode}

The primary comparisons leave a positive contribution from the latest
episode unestablished. Masking and UniSAR's matched replacement give
intervals that cross zero (Figure~\ref{fig:checkpoint-control-sensitivity}(a)).
Each intervention changes just the latest non-padding slot.

Wins and losses against the original episode nearly balance. The
preference intervals for UniSAR seeds 1 and 2 favor the replacement,
while seed 3 and SESRec include the 0.500 symmetry reference
(Table~\ref{tab:checkpoint-summary}(b)).

Comparisons on the same instances tell a similar story. On SESRec's
8{,}631 category-matchable instances from 2{,}134 users, the estimates
in $10^{-3}$ RR are $-0.183\,[-1.259,+0.835]$ for masking and
$-0.213\,[-1.135,+0.783]$ for replacement. The paired gap is
$+0.029\,[-1.112,+1.127]$. The mask interval is 95\%, and the
replacement and paired intervals are 97.5\%. On UniSAR's 87{,}044
matchable instances, the mask gives a negative estimate for seed 1
($-0.486\,[-0.931,-0.036]\times10^{-3}$ RR). The seed 2 and 3
intervals include zero.

\subsection{The Full Search History}

The full-history estimate depends on what replaces the original search.
All three mask intervals are above zero on the same matchable instances.
With matched replacement, the original search gives lower mean RR
for seed 1, and the intervals for seeds 2 and 3 include zero.
The paired gap is positive for every seed, from $+2.770\times10^{-3}$
to $+6.604\times10^{-3}$ RR (Figure~\ref{fig:checkpoint-control-sensitivity}(b--c)).
The tie-adjusted preference for the original history ranges from 0.4899
to 0.4959 (Table~\ref{tab:checkpoint-summary}(b)).

Replacing the full search history with matched searches changes the
target's rank in 44.1--44.8\% of instances, even though the original
history shows little average advantage over the replacements. Among the
ranks that change, 47.7--49.1\% worsen. Masking moves more ranks
(59.7--61.7\%), and worsening is more common among those changes
(53.3--55.3\%). When the rank moves, the median absolute shift is one
position with matched replacement and two with masking. The 90th
percentiles are four and nine to ten positions, respectively.

When we prioritize profile without requiring a category match, the
estimates range from $-1.600\times10^{-3}$ to $-0.068\times10^{-3}$ RR.
The seed 1 interval is below zero, and the other two cross zero. The
contrast with masking thus extends to a second replacement rule.

\subsection{Scope and Sensitivity}

We build the matched control by assigning observed search IDs to one
another. The map covers 45{,}895 IDs with content. Each mapped ID receives a
different donor ID, and each donor ID maps from only one original ID. Matching uses category and
profile buckets for item count, query length, and frequency. It
uses no user, target, candidate, label, rank, or model score. Masking
the full history gives positive contributions both among instances
with matches and among those without.

We varied the offsets within matching buckets to build five alternative
assignments before scoring them. Each assignment uses the same
checkpoints and instances, with 10{,}000 user-cluster resamples.
The 97.5\% intervals account for the two scopes within each
assignment and checkpoint. Masking attributes more benefit to the original
history in all fifteen comparisons, with paired gaps from $+2.331\times10^{-3}$
to $+7.106\times10^{-3}$ RR and all intervals above zero
(Table~\ref{tab:checkpoint-summary}(c)). Matched replacement again leaves
a positive contribution unestablished for the full history and for the
latest episode on the 87{,}044 matchable instances. The same users recur
across assignments.

The profile-prioritized rule changes the category mix. On the same
66{,}139 instances, 10.6\% of non-padding search slots share a top-level
category with their replacement. Across histories, the mean fraction of
original categories retained is 47.4\%. The original mask-versus-match
contrast also varies with history length. In the 4--30-slot strata, all
12 seed-by-stratum estimates are positive and 11 pointwise 97.5\%
intervals exclude zero. In the 1--3-slot strata, all six include zero.

\paragraph{Earlier searches and intervention scope.}
Replacing the latest search with the nearest earlier distinct search
from the same user (C3) gives a positive contribution for UniSAR seed 3
on 155{,}678 instances ($+0.238\,[+0.017,+0.464]\times10^{-3}$ RR,
pointwise 95\% CI). The other two intervals include zero. Full-history C3
rotates search IDs within each instance to change all slots. Its three
intervals include zero. We also intervene on the latest 2, 4, or 8 slots,
with 155{,}678, 141{,}172, or 113{,}555 eligible instances, respectively.
With pointwise 95\% intervals, masking eight slots gives a positive
contribution only for seed 1. Masking all slots gives positive
contributions at all three checkpoints. Profile-prioritized replacements
leave a positive contribution unestablished at these scopes.

\paragraph{Comparison on shared candidates.}

To compare the models on a shared evaluation set, we retrained both with
episode-level IDs and a transductive context vocabulary, then evaluated them
on the same 3{,}249 episodes with 100 candidates each. Full-input RR ranges
from 0.0637 to 0.1138 for SESRec and from 0.1705 to 0.1928 for UniSAR.
Both models receive a single search session in this setting. Masking that
session gives a contribution interval above zero in two of three SESRec runs
and none of the UniSAR runs. When we replace the session with a category-matched
search, all six intervals for preference based on the target item's score
include 0.500. Both comparisons use pointwise 95\% user-cluster confidence
intervals.

\section{Discussion}

\paragraph{Why the reference input matters.}
The two contribution estimates start from the same original prediction.
Their difference tells us how much higher RR is with the matched history
than with the mask. A matched search may retain category or activity
information the scorer uses, and a neutral ID may disturb a scorer that
rarely sees it. Either mechanism could account for part of the gap. The
mask measures what changes when search content is removed. The matched
comparison asks whether the original history improves ranking over
observed searches with similar measured properties.

\paragraph{What to report in a search audit.}
\begin{enumerate}
  \item \textbf{Scope.} Name the inputs that change, such as the latest
  episode or the full search history.
  \item \textbf{Reference.} Describe the mask or the matching rule. Compare
  controls on the same eligible instances and candidate sets when evaluating
  their difference.
  \item \textbf{Ranking results.} Report how often the target moves, the
  signed RR difference, and the preference for the original over matched
  searches. Give the uncertainty and eligible sample size.
\end{enumerate}

\section{Conclusion}

The primary comparisons in SESRec and UniSAR leave a ranking benefit from
the latest search episode unestablished. In UniSAR, estimates for the full
search history depend on the reference input, even though both masking
and matched replacement often move the target. FocalSwap can detect known
search signals in the planted and GenSAR calibrations and helps locate
where evidence for a benefit comes from. A search audit should report the
inputs it changes, the replacements it uses, and how the ranking responds.

\section{Reproducibility}

The \tool{} repository (\url{https://github.com/jdding/focalswap}) provides
the audit code, examples, and aggregate results, with scripts to regenerate
selected tables and figures. Rerunning the checkpoint evaluations requires
the original datasets, model implementations, and compatible checkpoints.

\section{Ethical Considerations}

We report aggregate analyses of public benchmark logs and do not
redistribute user-level data. Search data may reveal sensitive intent;
findings concern the audited datasets, checkpoints, and interfaces.

\bibliographystyle{ACM-Reference-Format}
\bibliography{references}

\end{document}